 \documentclass[prl,aps,twocolumn,showpacs]{revtex4}

\usepackage{epsfig}
\usepackage{amsfonts}
\usepackage{amsmath}

\begin{document}

\title{Local evidence for collective spin excitations in a distorted kagome antiferromagnet: Pr$_3$BWO$_9$}

\author{K. Y. Zeng$^{1,2}$}
\author{F. Y. Song$^{3}$}
\author{Z. M. Tian$^{3}$}
\email{tianzhaoming@hust.edu.cn}
\author{Qiao Chen$^{4}$}
\author{Shun Wang$^{4}$}
\author{Bo Liu$^{5,6}$}
\author{Shiliang Li$^{5,6}$}
\author{L. S. Ling$^{1}$}
\author{W. Tong$^{1}$}
\author{Long Ma$^{1}$}
\email{malong@hmfl.ac.cn}
\author{Li Pi$^{1,2}$}

\affiliation{$^{1}$Anhui Province Key Laboratory of Condensed Matter Physics at Extreme Conditions, High Magnetic Field Laboratory, Chinese Academy of Sciences, Hefei 230031, China\\
$^{2}$ Hefei National Laboratory for Physical Sciences at the Microscale, University of Science and Technology of China, Hefei 230026, China\\
$^{3}$ School of Physics and Wuhan National High Magnetic Field Center, Huazhong University of Science and Technology, Wuhan 430074, PR China\\
$^{4}$ School of Physics and MOE Key Laboratory of Fundamental Physical quantum Physics, PGMF, Huazhong University of Science and Technology, Wuhan 430074, China\\
$^{5}$ Beijing National Laboratory for Condensed Matter Physics, Institute of Physics, Chinese Academy of Sciences, Beijing 100190, China\\
$^{6}$ School of Physical Sciences, University of Chinese Academy of Sciences, Beijing 100190, China
}

\date{\today}

\begin{abstract}

  We report the local probe investigation of a frustrated antiferromagnet Pr$_3$BWO$_9$ with the distorted kagome lattice. Absence of magnetic order or spin freezing is indicated by the spectral analysis down to 0.3 K and specific heat measurements down to 0.09 K. The Knight shifts show an upturn behavior with the sample cooling down, which is further suppressed by external field. For the spin dynamics, gapped spin excitation is observed from the temperature dependence of spin-lattice relaxation rates, with the gap size proportional to the applied magnetic field intensity. Comparatively, an unexpected sharp peak is observed in the nuclear spin-spin relaxation rate data at $T^*\sim 4-5$ K. These results indicate an unconventional persistent fluctuating paramagnetic ground state with antiferromagnetic collective spin excitations in the strongly frustrated spin system.

\end{abstract}

\maketitle

Geometrical frustration has triggered enormous research interest in recent years, not only for the possible relationship with high-$T_c$ superconductivity but also the interesting ground states and exotic quantum excitations in its own right\cite{Balents_Nature_2010}. For the antiferromagnetically coupled spins located at triangles, the anti-parallel configuration favored by the nearest interaction cannot be simultaneously fulfilled, leading to very large degeneracy of ground states as well as strong quantum fluctuations. One of the most intriguing ground state is the quantum spin liquid (QSL)\cite{Balents_Nature_2010}, where the spins dynamically entangle with each other but never order even at zero temperature. Among various types of QSL proposed by theory, the common feature of telling QSL from other states is the deconfined spinon excitations with fractional quantum number. As a result, research into the low-energy spin excitation property is of great importance in the study of frustrated antiferromagnets.

Compared with the edge-shared triangular lattice, the corner-shared triangles (kagome lattice) with low coordination and weak second-neighbor interaction possess stronger magnetic frustration effect, thus is more attractive. In the near decade, ZnCu$_3$(OH)$_6$Cl$_2$ is the intensively studied promising kagome lattice realizing the QSL state\cite{Norman_RMP_2016}, where observed is the continuous spin excitation\cite{Han_Nature_492_406}, fingerprints for the fractional spinons\cite{Balents_Nature_2010}. However, two inherent drawbacks exist in ZnCu$_3$(OH)$_6$Cl$_2$. One is the natural mixing of Cu$^{2+}$ and Zn$^{2+}$ due to their similar ion radius\cite{Imai_PRB_84_020411}, complicating the investigation of the intrinsic properties of QSL state. The other one is the almost impossibility of chemical substitution at the magnetic sites, prohibiting further exploration of other novel states and excitations in this system.

For magnetic lanthanide ions, the local crystal electric field (CEF) splits the spin-orbit entangled $J$ momentum space into $2J+1$ states, and finally determine the single ion anisotropy with effective $J_{eff}$. By locating different lanthanide ions to kagome lattice, various novel ground states and spin excitations can be realized, relying on different spin anisotropy, spin-orbit coupling and exchange and dipolar interaction. The rare-earth based Ln$_3$M$_2$Sb$_3$O$_{14}$ (Ln= lanthanide, M= Mg or Zn) with kagome lattice is a precise example for testing this judgement\cite{Dun_PRL_116_157201}. For Dy$_3$Mg$_2$Sb$_3$O$_{14}$ with anisotropic Ising spins, the kagome spin ice with "two-in-one-out" or "one-in-two-out" magnetic structure is identified by neutron scattering\cite{Paddison_NC_7_13842}. While for the Ln=Ho case, a new quantum state with both characteristics of classical spin ice and quantum fluctuation and tunneling is found\cite{Dun_PRX_10_031069}. For the Ln=Gd case with isotropic large Heisenberg large spins, the dipolar interaction may result in the observed magnetic ordered state with 120 degree structure\cite{Dun_PRL_116_157201,Wellm_PRB_102_214414}. Recently, possible QSL state is proposed  based on the persistent strong low-energy spin excitations down to $T=50$ mK by muon spin rotation ($\mu$SR) and the magnetic contribution to the linear temperature dependence of heat capacity\cite{Ding_PRB_98_174404}. Controversially, the random mixing of Tm$^{3+}$ and Zn$^{2+}$ is observed very recently\cite{Ma_PRB_102_224415}, which may lead to the mimicry of QSL. Thus, the rare-earth based kagome lattice has supplied a new playground for exploring exotic frustrated quantum states.

To overcome the possible anti-site disorder, some of us have synthesized a new family of rare-earth based antiferromagnets Ln$_3$BWO$_9$ (Ln=Pr, Nd, Gd-Ho) with the lanthanide ions locating on the distorted kagome structure in $ab$-plane and stacking in an $AB$-type fashion along $c$-axis\cite{Ashtar_IC_59_5368}. $Dc$ magnetization analysis suggests similar magnetic behaviors with other rare-earth based frustrated antiferromagnets\cite{Ashtar_IC_59_5368}. However, no further study on the ground state and spin excitations is reported up to now.

In this article, we employ nuclear magnetic resonance (NMR) as a local probe to study the spin excitations in the lanthanide frustrated antiferromagnet Pr$_3$BWO$_9$ with a distorted kagome lattice. The paramagnetic state persist without any magnetic ordering or spin freezing as evidenced by $^{11}$B spectral analysis down to $T=0.3$ K and specific heat measurements down to $T=0.09$ K, far below its Curie-Weiss temperature. The Knight shifts measuring spin correlations at $\mathbf{q}=0$ show an upturn behavior with the sample cooling down, which is further suppressed by external field. For the spin dynamics, thermally activated behavior is observed in the temperature dependence of spin-lattice relaxation rates, indicating gapped spin excitations. The gap size is proportional to the field intensity for both field directions, but with a very different slope, demonstrating the spin anisotropy. Comparatively, an unexpected sharp peak is observed in the nuclear spin-spin relaxation rate data at $T^*\sim 4-5$ K, which is contributed by the hyperfine field fluctuation parallel to the applied magnetic direction. These results indicate a cooperative paramagnetic state with short-ranged collective excitations in the strongly frustrated spin system.

Single crystals of Pr$_3$BWO$_9$ were synthesized by the conventional flux method. For NMR studies, crystals with typical dimensions of $1.5\times1.5\times0.2$ mm$^3$ are used. Our NMR measurements are conducted on $^{11}$B nuclei ($\gamma_n=13.655$ MHz/T, $I=3/2$) with a phase-coherent NMR spectrometer. The spectra are obtained by summing up or integrating the spectral weight by sweeping the frequency under fixed magnetic fields. The measurement of spin-lattice relaxation rates is performed by the inversion-recovery method. The spin-spin relaxation rates are obtained by measuring the nuclear transverse dephasing with the two-pulse Hahn echo sequence.

\begin{figure}
\includegraphics[width=8cm, height=8cm]{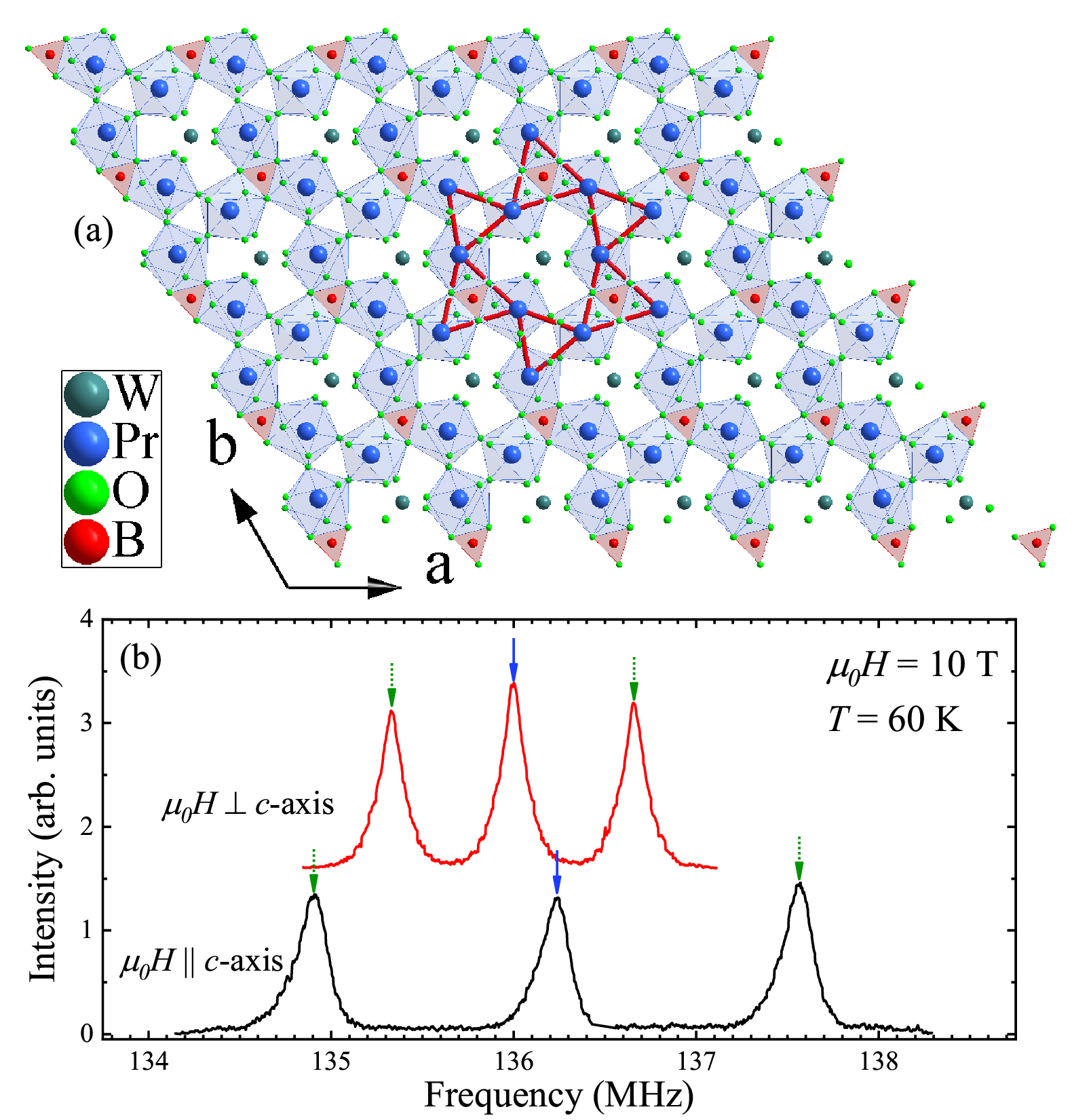}
\caption{\label{struc1}(color online) (a) The crystalline structure of Pr$_3$BWO$_9$ as seen against $c$-axis. The distorted kagome lattice is marked by the bold red lines. (b) Typical $^{11}$B NMR spectra with the magnetic field applied perpendicular or parallel to the $c$-axis of the single crystal. The central and satellite transitions are respectively marked by the solid blue and dotted green arrows.
}
\end{figure}

The crystal structure of Pr$_3$BWO$_9$ is built up by blocks of PrO$_8$, WO$_6$ and BO$_3$ polyhedra in the corner- or edge-sharing manner\cite{Ashtar_IC_59_5368}. There is only one Wyckoff position for Pr, W, and B sites. The BO$_3$ trigons share edges with the PrO$_8$ polyhedra, leading to a very strong hyperfine coupling between $^{11}$B nuclei and magnetic moments. Top view of the crystal structure along $c$-axis is shown in Fig.\ref{struc1}(a). The magnetic Pr$^{3+}$ sites form a distorted kagome structure in the crystalline $ab$-plane, which is intensionally marked with bold red lines.

In Fig.\ref{struc1} (b), we show typical $^{11}$B NMR spectra at $T=60$ K with a $10$ tesla field along different directions. Both spectrum are composed by three peaks, one at the frequency center and the other two locate symmetrically at both sides. For nuclei with spin $I=3/2$ in non-zero local electric field gradient (EFG) under strong magnetic fields, the first order correction to the Zeeman energy term due to quadruple interactions splits the single NMR transition into three, respectively named as the central transition and satellites\cite{Slichter_NMR,Abragam_book}. The frequency of satellites strongly depend on the angle $\theta$ between applied field and the pricipal axis of the EFG tensor. For the present sample without $ab$-plane EFG anisotropy (confirmed by in-plane angle rotation, data not shown), the correction to the frequency can be written as $\nu_m^{(1)}=\nu_Q(m-1/2)(3\cos^2\theta-1)/2$, where $\nu_Q$ and $m$ respectively denotes the quadruple frequency and nuclear magnetic quantum number. The $\nu_Q$ is calculated to be $\sim1.325$ MHz at $T=60$ K, and the main axis of EFG is along the crystalline $c$-axis. These observations are fully consistent with the crystal structure symmetry.

\begin{figure}
\includegraphics[width=8cm, height=9cm]{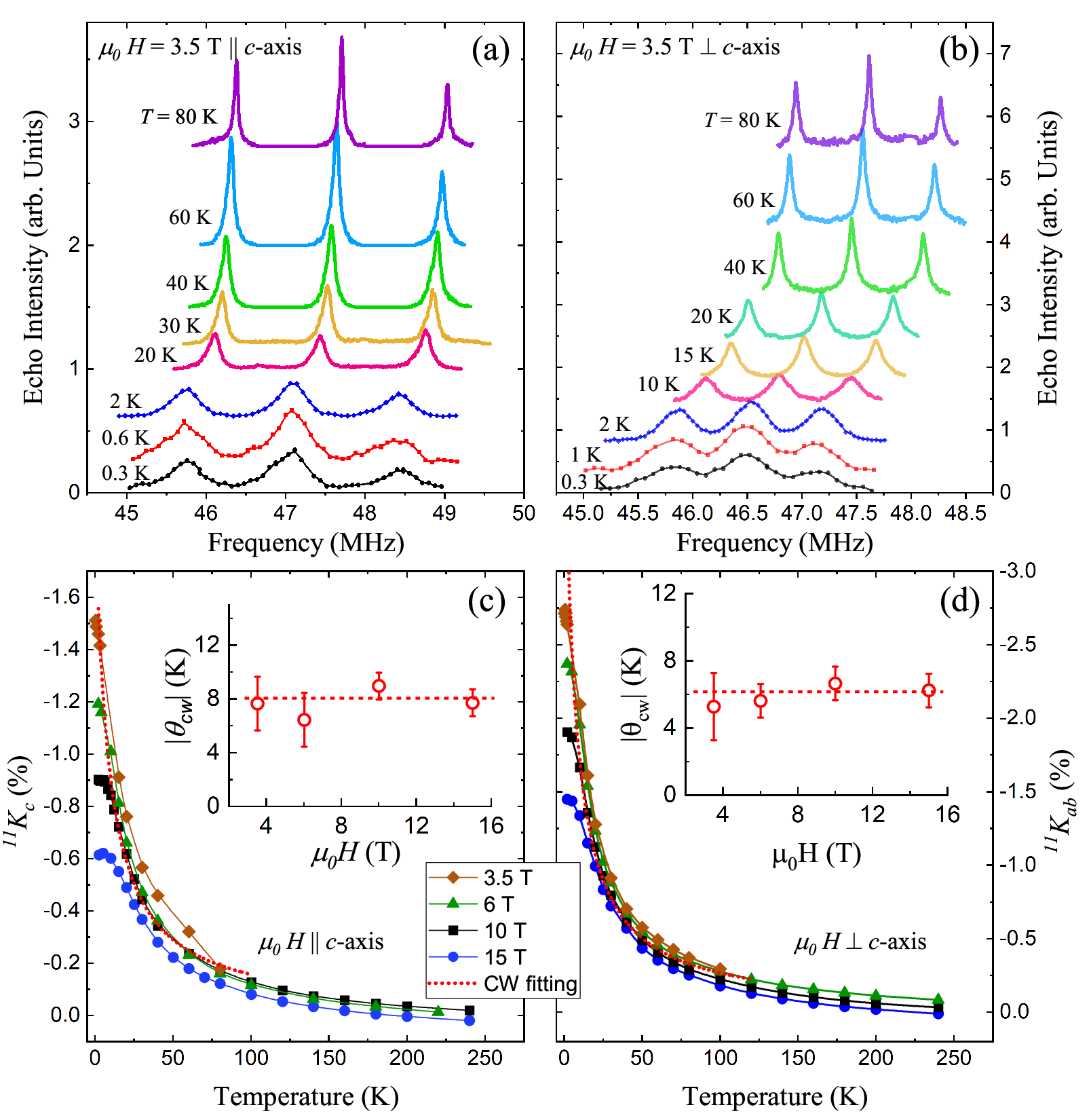}
\caption{\label{knight2}(color online)
The frequency-swept $^{11}$B NMR spectra at different temperatures with a 3.5 Tesla field applied along (a) or perpendicular to $c$-axis(b).
(c) and (d): The temperature dependence of $^{11}$B Knight shift $^{11}K$ under different magnetic fields. The dotted lines are fits to the Curie-Weiss function. The absolute value of obtained Curie-Weiss temperatures ($|\theta_{cw}|$) as a function of field intensity is plotted in the inset.
}
\end{figure}

The full NMR spectra at different temperatures with the magnetic field applied parallel or perpendicular to the $c$-axis are shown in Fig.\ref{knight2} (a) and (b). For both field directions, all the three peaks shift to lower frequency side and broaden gradually with the sample cooling down. For $T=0.3$ K far below its Curie-Weiss temperature ($|\theta_{cw}|=6.1$ K for $\mu_0H\perp c$ and $|\theta_{cw}|=5.4$ K for $\mu_0H|| c$, as shown by the single crystal dc-susceptibility\cite{SM}),the main feature of the spectrum is maintained as that at much higher temperatures.
No further line splitting or amplitude modulation is observed in the present sample. We note that the hyperfine coupling between $^{11}B$ nuclei and the magnetic sites is very strong as evidenced by the enormous Knight shift and ultra-fast spin-lattice relaxation (shown latter). Any magnetic ordering of commensurate or incommensurate type or spin freezing, if present, is very unlikely to be missed in the low-temperature spectra for both field directions. Additionally, no magnetic transition is observed from the specific heat measurements down to 0.09 K\cite{SM}.
Thus, our data support a paramagnetic state in Pr$_3$BWO$_9$ with the temperature down to at least $0.017|\theta_{cw}|$, demonstrating the strong magnetic frustration.

The intrinsic local spin susceptibility can be measured by the Knight shift, suffering less from impurity effect than bulk probes. The temperature dependence of Knight shift for different magnetic fields is shown in Fig.\ref{knight2} (c) and (d). The Knight shifts ($^{11}K$) are calculated as the relative line shift of the central transition with respect to the Larmor frequency, after removing the second order correction arising from quadruple interaction. With the sample cooling down, all the Knight shifts share the similar Curie-Weiss upturn behavior with $dc$ susceptibility. Fittings with the equation $^{11}K \propto 1/(T-\theta_{cw})$ to the low temperature data yield a nearly field-independent $|\theta_{cw}|$ value (shown in the inset). The $\theta_{cw}$ is calculated to be $\sim-6$ K for $\mu_0H\perp c$ and $\sim-8$ K for $\mu_0H|| c$, again demonstrating a moderate antiferromagnetic coupling strength. With the increasing field intensity, the Knight shift tends to level off at low temperatures. This results from the saturated paramagnetic Pr$^{3+}$ magnetic moments, which is also observed in other frustrated\cite{Zeng_PRB_102_045149,Baenitz_PRB_98_220409,Bordelon_NP_15_1058} or Kitaev antiferromagnets\cite{Zheng_PRL_119_227208}.

\begin{figure}
\includegraphics[width=8cm, height=7cm]{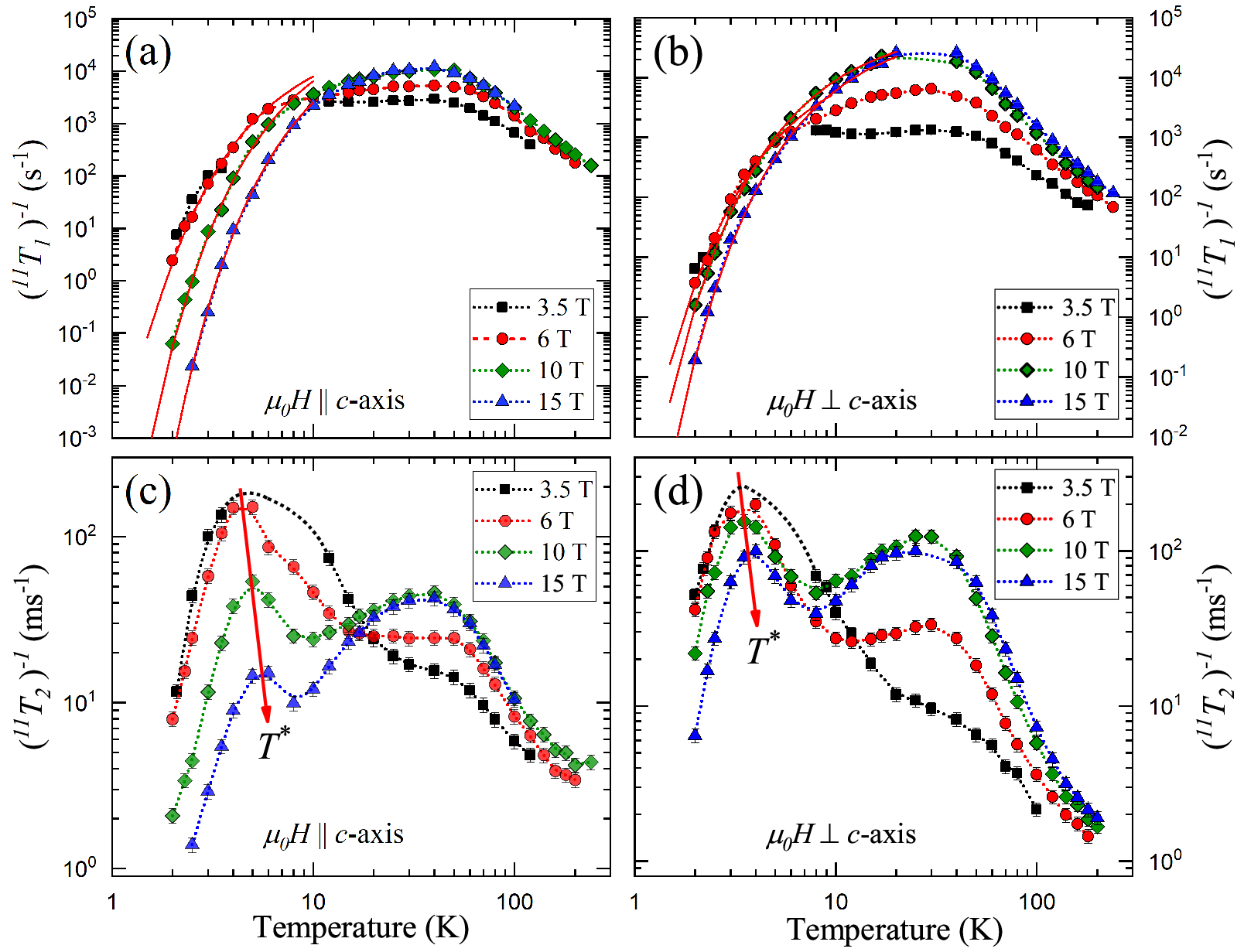}
\caption{\label{relax3}(color online)
  (a)(b): Spin-lattice relaxation rates ($1/^{11}T_1$) as a function of temperature under different magnetic fields with $\mu_0H || c$ or $\mu_0H\perp c$. The solid red lines are representative fits to $1/T_1\propto \exp(-\Delta/T)$ (See the text).
  (c)(d): Spin-spin relaxation rates ($1/^{11}T_2$) as a function of temperature under different magnetic fields with $\mu_0H||c$ or $\mu_0H\perp c$.
}
\end{figure}

To further explore the spin dynamics in Pr$_3$BWO$_9$, we have measured both nuclear spin-lattice and spin-spin relaxation rates ($1/^{11}T_1$ and $1/^{11}T_2$), and show their temperature dependence under different magnetic fields in Fig.\ref{relax3}. Both measurements are made at the central peak. The $1/^{11}T_1$ is obtained by fitting the nuclear magnetization to the function, $M(t)/M(\infty)= 1-0.1\exp(-t/T_1)-0.9\exp(-6t/T_1)$, the standard recovery function for the central transition of nuclei with $I=3/2$ locating at non-zero local electric field gradient(EFG)\cite{Narath_PR_162_320}. An additional stretching factor $\beta$ is needed for temperatures below $T\sim40$ K for the observable distribution of the relaxation. For $1/^{11}T_2$, the echo intensity decay can be well reproduced by the function, $M(2\tau)=M(0)\exp(-2\tau/T_2)$ ($\tau$ is the time interval between $\pi/2$ and $\pi$ pulse), for the studied temperature range. This demonstrates the transverse dephasing is mainly contributed by the spin fluctuations via Redfield mechanism\cite{Redfield_PR_98_1787} instead of indirect and dipole coupling between nuclei, which will result in a gaussian component\cite{Takigawa_PRB_57_1124}. For the high temperature region of $T>20$ K, the NMR relaxation rates $1/^{11}T_1$ and $1/^{11}T_2$ share a very similar temperature dependence, first increase mildly, then tend to level off and finally bend over with the sample cooling down. A wide hump peaked at $T\sim30$ K is observed in both relaxation rates.

The high temperature hump is contributed by the crystal electric field (CEF)-related excitations. The magnetic behavior of rare-earth ions strongly depends on the local CEF where they occupy. The $(2J+1)$-fold degenerated states of $4f$-electrons are lifted by a certain CEF. For the non-Kramers ion Pr$^{3+}$with $J=4$, the crystal field interactions lead to nine singlet states with possible accidental- or near-degeneracies\cite{Scheie_PRB_98_134401}. The mild increasing $1/^{11}T_1$ and $1/^{11}T_2$ at high temperatures directly connect with the slowing down of spin fluctuations of electrons occupying thermally activated CEF levels\cite{Lumata_PRB_81_224416,Zeng_PRB_102_045149}. The wide hump results from the reduced thermal population at low temperatures, consistent with the slope change observed in $dc$-susceptibility\cite{Ashtar_IC_59_5368}.

\begin{figure}
\includegraphics[width=8cm, height=10cm]{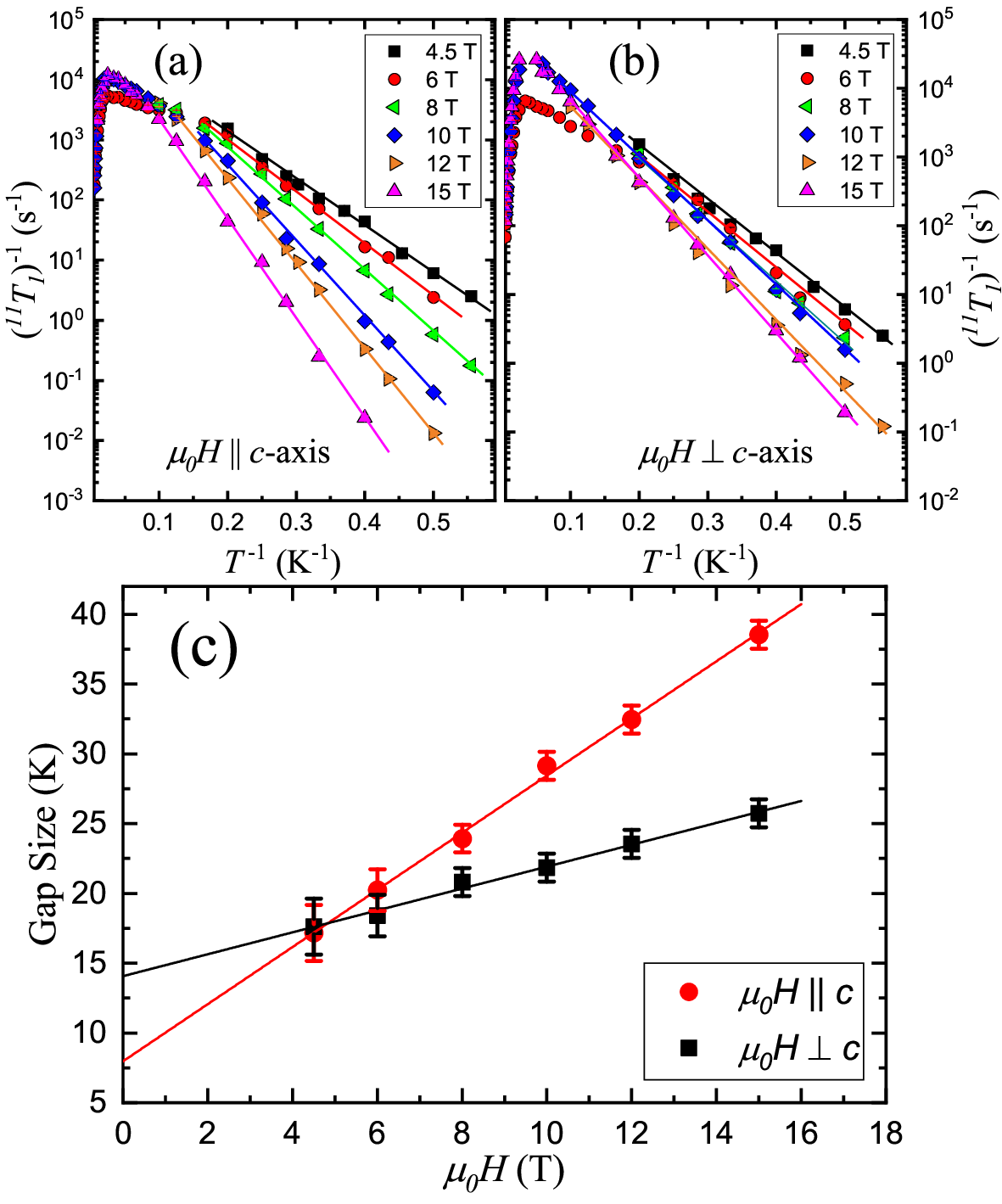}
\caption{\label{gap4}(color online)
  The Arrhenius plot of $1/^{11}T_1$ versus inverse temperature $1/T$ for magnetic fields along (a) or perpendicular (b) to $c$-axis. Solid lines are linear fits to the data.
  (c): Field dependence of the spin excitation gap in Kelvin. Solid lines are fits to $\Delta(H)=\Delta_0+g\mu_BH$.
}
\end{figure}

More intriguing is the low temperature spin dynamics dominated by the CEF ground state of Pr$^{3+}$. With the sample further cooling from $T=10$ K, the $1/^{11}T_1$ drops steeply with temperature for all the applied fields (See Fig.\ref{relax3}(a) and (b)). Actually, the temperature dependence of $1/^{11}T_1$ can be well described by the thermally activated equation $1/T_1\propto\exp(-\Delta/T)$, with a finite spin gap $\Delta$. The gapped spin excitations under magnetic fields are better demonstrated by the Arrhenius plot, i.e., $1/^{11}T_1$ versus inverse temperature on a semilogarithmic scale as shown in Fig.\ref{gap4} (a) and (b). The field dependence of the gap size is shown in Fig.\ref{gap4}(c) for both field orientations. In sharp contrast, the temperature dependence of $1/^{11}T_2$ at $T<10$ K completely deviates from that of $1/^{11}T_1$. A prominent peak around $T^*\sim4$ K is observed, which is unusual in strongly correlated electron systems. When the magnetic field strength increases, the peak is suppressed, and shifts to a little higher temperature, which is more pronounced for fields along $c$-axis.

We discuss the spin excitation property in Pr$_3$BWO$_9$ implied by our data. The low-temperature Knight shifts show a Curie-Weiss upturn behavior, while a contrasting gapped behavior is observed in $1/^{11}T_1$ with the sample cooling down. By roughly extrapolating the field dependence of the gap size (See Fig.\ref{gap4}(c)), a non-zero spin gap should also exist at zero field. In another word, this excitation gap is not field-induced, but inherent. The Knight shift,expressed as $K\propto A_{hf}\chi(\mathbf{q}=0)$, measures the spin excitation at the wave-vector $\mathbf{q}=0$\cite{Slichter_NMR,Abragam_book}. The hyperfine coupling constant $A_{hf}$ is negative in the present sample. However, the $1/^{11}T_1$ is contributed by the dynamic spin susceptibility summed over the whole reciprocal $\mathbf{q}$-space, which can be formulated as $1/T_1\propto T\sum_{\mathbf{q}}|A(\mathbf{q})|^2[\chi^{''}(\mathbf{q},\omega_L)]/\omega_L$\cite{Slichter_NMR,Abragam_book}. The $\omega_L$ denotes the Larmor frequency, which can be treated as zero in condensed matter physics. Thus, the contrasting low temperature behavior of $1/^{11}T_1$ indicates the strongly $\mathbf{q}$-dependent spin excitation, and the gap opens at a non-zero wave-vector. In such an antiferromagnet with moderate coupling strength, this result actually suggests the spin excitation gaps at the antiferromagnetic wave-vector.

The obvious $\mathbf{q}$-dependent spin excitations in Pr$_3$BWO$_9$ is very different with the mostly investigated spin liquid candidate, ZnCu$_3$(OH)$_6$Cl$_2$. Fractional excitation continuum is observed by inelastic neutron scattering in the later compound\cite{Han_Nature_492_406}. By the recent NMR results\cite{Fu_Science_350_655}, the spin gapped behavior is observed in both Knight shifts and spin-lattice relaxations, which further indicates a gapped spin liquid state in ZnCu$_3$(OH)$_6$Cl$_2$. Additionally, the spin gap is suppressed by the applied magnetic field due to Zeeman effect, whose slope corresponds to spinons with $S=1/2$. Similar results are observed in its counterpart Cu$_3$Zn(OH)$_6$FBr\cite{Feng_CPL_34_077502}. Suppose spin liquid state do exist in these materials, the spin gap originates from the majority of short-ranged valence bonds.

There exist short-ranged collective spin excitations in the present sample. In contrast to the gapped behavior observed in low temperature $1/^{11}T_1$, a sharp peak is observed in the nuclear spin-spin relaxation rate ($1/^{11}T_2$) at $T^*\sim4$ K. The spin fluctuation contribution to ($1/^{11}T_2$) through the Redfield mechanism can be expressed as $1/T_2^{Redfield}=1/T_2'+1/(2T_1)$\cite{Slichter_NMR}. The first term describes the in-plane decoherence contributed by the fluctuations of longitudinal hyperfine field. The later one has the same origination as  $1/^{11}T_1$, both from the transverse hyperfine field fluctuations. Thus, the different behavior at low temperatures apparently results from the fluctuating hyperfine field anisotropy.

Understanding how longitudinal fluctuations contribute $1/T_2'$ supplies the key ingredient for revealing the spin excitation property. In the two-pulse Hahn echo experiment, dephasing of the transverse nuclear magnetization during the timing interval between pulses is refocused by applying the second $\pi$-pulse after the equal time interval. If there exists some kinds of dynamically inhomogeneous longitudinal field with the typical frequency of kilohertz, the changed local magnetic field will give rise to the failure of refocusing, also the decreased echo intensity. One of typical examples for this effect is the well-known vortex in type-II superconductors under magnetic field\cite{Suh_PRL_71_3011}. Measurements of $1/T_2$ have supplied another novel approach to study the vortex dynamics in both cuprates and iron based high-$T_C$ superconductors\cite{Recchia_PRL_78_3543,Bossoni_PRB_85_104525,Rigamonti_RPP_61_1367}. The vortex core movement contributes a very similar enhanced relaxation as observed here. Another example is fluids in a pore space. The spin-spin relaxation is obviously faster than the pure fluids, which is enhanced by the interactions between nuclei and pore walls or paramagnetic centers\cite{Korringa_PR_127_1143}. Thus, the enhanced $1/T_2$ should be related with excitations of typical collective characteristic.

In other distorted kagome systems Nd$_3$Ga$_5$SiO$_{14}$ and Pr$_3$Ga$_5$SiO$_{14}$, electron spin resonance study have revealed the collective spin excitations instead of the long-range continuum\cite{Ghosh_PRB_88_094414,Ghosh_PRB_90_224405}. In Pr$_3$Ga$_5$SiO$_{14}$, Dynamical short-range ordering is suggested from diffuse neutron-scattering observations\cite{Lumata_PRB_81_224416} . $\mu$SR experiments indicate a fluctuating collective paramagnetic ground state in Nd$_3$Ga$_5$SiO$_{14}$\cite{Zorko_PRL_100_147201}. From the NMR experiments, a wipeout of the $^{71}$Ga NMR signal below $T=25$ K is observed in Nd$_3$Ga$_5$SiO$_{14}$\cite{Zorko_PRL_100_147201}, resulting from the enhanced $1/T_2$. In Pr$_3$Ga$_5$SiO$_{14}$, The enhanced $1/T_2$ at low temperatures is also observed by NMR\cite{Lumata_PRB_81_224416}, although the physical connection between them is not pointed out. In the present Pr$_3$BWO$_9$, our observations provide unambiguous local evidence for the short-ranged collective excitations. We stress that the short-range correlations still is dynamic in the timescale of NMR for the studied magnetic field range, as no abrupt change is observed in the temperature dependence of NMR spectra. This is in sharp contrast with what occurs in Nd$_3$Ga$_5$SiO$_{14}$, where field-induced partial magnetic order is identified by neutron scattering and heat transport\cite{Zhou_PRL_99_236401,Li_PRB_85_174438}.

The field dependence of the spin excitation gap can be understood with a simple theoretical model. Based on the assumption that the Pr$^{3+}$ also possess the $S=1$ spin state as what occurs in Pr$_3$Ga$_5$SiO$_{14}$\cite{Bordet_JPCM_18_5147}, the Hamiltonian of the spin system only considering the single ion anisotropy and antiferromagnetic exchange coupling can be roughly written as\cite{Ghosh_PRB_88_094414},
\[
\begin{split}
H=D\sum_i(S_i^z)^2+g\mu_BB\sum_iS_i^z+\\
J\sum_{i,j}[S_i^zS_j^z+\frac{1}{2}(S_i^+S_j^-+S_i^-S_j^+)].
\end{split}
\]
The $D$ and $J>0$ respectively denote the single iron anisotropy and antiferromagnetic exchange coupling strength. Ignoring the magnetic frustration effect is believed not to change the qualitative main predictions\cite{Ghosh_PRB_88_094414}. For the short-ranged excitation, results can be reached for two cases based on linearized spin-wave approximation,
(1) For $D>nJ$ ($n$ is the number of the nearest neighbors), the excitation energy spectrum is given by $\Delta E(\overrightarrow{k})=D\mp g\mu_BB+nJ\gamma_{\overrightarrow{k}}$, where $\gamma_{\overrightarrow{k}}$ is the structure factor.
(2) For $D<0$, the excitation energy is $\Delta E(\overrightarrow{k})=(nJ+|D|)\sqrt{1-(\lambda\gamma_{\overrightarrow{k}})^2}+g\mu_BB$.
For both cases, the spin excitation gap $\Delta$ at zero field is given by $\Delta(0)\sim nJ+|D|$. As predicted by case (1), the spin excitation gap will first decrease to zero and then increase linearly with applied magnetic field. While, the the gap size will only increase linearly with the field intensity for the $D<0$ case.

In Fig.\ref{gap4}(c), we show the field dependence of the spin excitation gap in the present Pr$_3$BWO$_9$ sample. The linear fittings to the monotonically increasing gap size yield $\Delta(0)= 8.0$ K and $g=3.047$ for $\mu_0H||c$-axis, and $\Delta(0)= 14.1$ K and $g=1.167$ for $\mu_0H\perp c$-axis. The monotonic field dependence of the gap size, positive intercept as well as much larger $g$ factor along the crystalline $c$-axis all support the $D<0$ case, where the $S^z=\pm1$ state is energetically favorable. The comparatively large $\Delta(0)$ with Curie-Weiss temperature should result from the pronounced single ion anisotropy.

The single ion anisotropy (determined by $D$) can be further qualitatively understood by the spin-orbit coupling effects ($\hat{H}_{SOC}=\lambda\hat{\mathbf{S}}\cdot\hat{\mathbf{L}}$) in perturbation theory\cite{Xiang_DT_42_823,Whangbo_ACR_48_3080}. The spin orientation preference is determined by the local energy state of PrO$_8$ polyhedra, i.e., the combination of spin states, highest occupied molecular orbital (HOMO) and lowest unoccupied molecular orbital (LUMO). For HOMO and LUMO with the same spin state, the spins prefer $||z$-direction when the change of orbital magnetic quantum number $|\Delta L_z|=0$, but prefer $\perp z$-direction when $|\Delta L_z|=1$. For HOMO and LOMO with different spin states, the situation for $|\Delta L_z|=0$ and $|\Delta L_z|=1$ is reversed.
Thus, it is essential to further identify the CEF splittings of the PrO$_8$ polyhedra and perform more strict theoretical simulation as that in GdInO$_3$\cite{Yin_arxiv_2106_05071}.

The ground state of Pr$_3$BWO$_9$ is neither magnetic ordered nor quantum spin liquid state at least for the present temperature range. The collective spin excitations based on spin clusters or loops exist in the sample, which is still dynamic for the studied temperature and field range. Strong $\mathbf{q}$-dependent behavior of the spin excitations is unambiguously evidenced by the contrasting temperature dependence of Knight shifts and spin-lattice relaxation rates. These are distinctively different with the long-range continuum excitations in quantum spin liquids. Within these constraints, the Pr$_3$BWO$_9$ may be regarded as cooperative paramagnet\cite{Villain_ZPB_33_31}.

In conclusion, we have performed detailed NMR study on the newly-synthesized distorted kagome spin system Pr$_3$BWO$_9$. Cooperative paramagnetic state persists down to $T=0.09$ K far below its Curie-Weiss temperature, yielding strong frustration effect. The unconventional short-ranged collective spin excitations is identified by comparatively studying the temperature dependence of Knight shifts, spin-lattice relaxation rates and spin-spin relaxation rates. A spin gap opened at the antiferromagnetic wave-vector is suggested from our data, whose size further shows linear field dependence. This short-ranged collective spin excitations with a distinct $\mathbf{q}$-dependent spin gap is totally different from the short-ranged valence bonds proposed in the spin liquid state. Our experiments reveal a highly unconventional spin excitation mode in frustrated antiferromagnets. Moreover, we have supplied a new approach to detect collective spin excitations in magnetic materials.

This research was supported by the National Key Research and Development Program of China (Grant No. 2016YFA0401802), the National Natural Science Foundation of China (Grants No. 11874057, 11874158, U1732273 and 21927814) and the Users with Excellence Program of Hefei Science center CAS (Grant No. 2019HSC-UE008). A portion of this work was supported by the High Magnetic Field Laboratory of Anhui Province.


\end{document}